\documentclass[11pt,letterpaper,fleqn]{article}
\usepackage{graphicx}
\usepackage{lineno}
\usepackage{hyperref}
\usepackage{afterpage}
\usepackage{pdflscape}
\hypersetup{
  colorlinks=true,
  urlcolor=blue,
  citecolor=blue
}
\usepackage{subfigure}
\usepackage{cite}
\usepackage{xcolor,colortbl}
\usepackage{wrapfig}
\usepackage{enumitem} 

\selectcolormodel{rgb}
\definecolor{LHCb dark}{rgb}{0.0000,0.3412,0.6549}
\definecolor{UC red}{rgb}{0.8196,0.1176,0.2314} 
\definecolor{brickred}{rgb}{0.8, 0.25, 0.33}
\definecolor{Gray}{gray}{0.85}

\let\oldenumerate\itemize
\renewcommand{\itemize}{
  \oldenumerate
  \setlength{\itemsep}{1pt}
  \setlength{\parskip}{1pt}
  \setlength{\parsep}{1pt}
}

\def\s2i2{$S^2 I^2$}

\begin{document}
\thispagestyle{empty}

\begin{flushright}
HSF-CWP-2017-16
\end{flushright}

\Large
\begin{center}
{\bf HEP Community White Paper on Software trigger\\ and event reconstruction}
\end{center}
\vskip 1cm

\normalsize

\hangindent=1cm
{\bf Abstract:} Realizing the physics programs of the planned and upgraded high-energy physics (HEP) experiments over the next 10 years will require the HEP community to address a number of challenges in the area of software and computing. For this reason, the HEP software community has engaged in a planning process over the past two years, with the objective of identifying and prioritizing the research and development required to enable the next generation of HEP detectors to fulfill their full physics potential. The aim is to produce a Community White Paper (CWP)~\cite{HSF2017} which will describe the community strategy and a roadmap for software and computing research and development in HEP for the 2020s. The topics of event reconstruction and software triggers were considered by a joint working group and are summarized together in this document.

\vskip 1cm

\hangindent=1cm
{\bf Editors}: Vladimir Vava Gligorov$^{12,a}$ and David Lange$^{20}$

\vskip 0.2cm
\hangindent=1cm
{\bf Contributors}:
Albrecht, Johannes$^{1}$;
Bloom, Kenneth$^{2}$;
Boccali, Tommaso$^{3}$;
Boveia, Antonio$^{4,c}$;
De Cian, Michel$^{5}$;
Doglioni, Caterina$^{6,b}$;
Dziurda, Agnieszka$^{7}$;
Farbin, Amir$^{8}$;
Fitzpatrick, Conor$^{9}$;
Gaede, Frank$^{10}$;
George, Simon$^{11}$;
Gligorov, Vladimir$^{12,a}$;
Grasland, Hadrien$^{13}$;
Grillo, Lucia$^{14}$;
Hegner, Benedikt$^{7}$;
Kalderon, William$^{6}$;
Kama, Sami$^{15}$;
Koppenburg, Patrick$^{16}$;
Krutelyov, Slava$^{17}$;
Kutschke, Rob$^{18}$;
Lampl, Walter$^{19}$;
Lange, David$^{20}$;
Moyse, Ed$^{21}$;
Norman, Andrew$^{18}$;
Petric, Marko$^{7}$;
Polci, Francesco$^{12}$;
Potamianos, Karolos$^{10}$;
Ratnikov, Fedor$^{22}$;
Raven, Gerhard$^{16}$;
Ritter, Martin$^{23}$;
Rizzi, Andrea$^{3}$;
Rodrigues, Eduardo$^{24}$;
Rousseau, David$^{13}$;
Salzburger, Andy$^{7}$;
Sexton Kennedy, Liz$^{18}$;
Sokoloff, Michael D$^{24}$;
Stewart, Graeme$^{7}$;
Ustyuzhanin, Andrey$^{22}$;
Viren, Brett$^{25}$;
Williams, Mike$^{26}$;
Winklmeier, Frank$^{27}$;
Wuerthwein, Frank$^{17}$
\bigskip 
\par {\footnotesize $^{1}$ Technische Universit\"at Dortmund, Dortmund, Germany}
\par {\footnotesize $^{2}$ University of Nebraska-Lincoln, Lincoln, NE, USA}
\par {\footnotesize $^{3}$ INFN Sezione di Pisa, Università di Pisa, Scuola Normale Superiore di Pisa, Pisa, Italy}
\par {\footnotesize $^{4}$ The Ohio State University, Columbus, OH, USA}
\par {\footnotesize $^{5}$ Physikalisches Institut, Ruprecht-Karls-Universitat Heidelberg, Heidelberg, Germany}
\par {\footnotesize $^{6}$ Fysiska institutionen, Lunds Universitet, Lund, Sweden}
\par {\footnotesize $^{7}$ CERN, Geneva, Switzerland}
\par {\footnotesize $^{8}$ Department of Physics, The University of Texas at Arlington, Arlington, TX, USA}
\par {\footnotesize $^{9}$ Institute of Physics, École Polytechnique F\'ed\'erale de Lausanne (EPFL), Lausanne, Switzerland}
\par {\footnotesize $^{10}$ Deutsches Elektronen-Synchrotron, Hamburg, Germany}
\par {\footnotesize $^{11}$ Department of Physics, Royal Holloway University of London, Surrey, United Kingdom}
\par {\footnotesize $^{12}$ LPNHE, Université Pierre et Marie Curie, Université Paris Diderot, CNRS/IN2P3, Paris, France}
\par {\footnotesize $^{13}$ LAL, Université Paris-Sud and CNRS/IN2P3, Orsay, France}
\par {\footnotesize $^{14}$ School of Physics and Astronomy, University of Manchester, Manchester, United Kingdom}
\par {\footnotesize $^{15}$ Physics Department, Southern Methodist University, Dallas TX, United States of America}
\par {\footnotesize $^{16}$ Nikhef National Institute for Subatomic Physics and Vrije Universiteit Amsterdam, Amsterdam, The Netherlands}
\par {\footnotesize $^{17}$ University of California, San Diego, La Jolla, CA, USA}
\par {\footnotesize $^{18}$ Fermi National Accelerator Laboratory, Batavia, IL, USA}
\par {\footnotesize $^{19}$ Department of Physics, University of Arizona, Tucson, AZ, USA}
\par {\footnotesize $^{20}$ Princeton University, Princeton, NJ, USA}
\par {\footnotesize $^{21}$ Department of Physics, University of Massachusetts, Amherst, MA, USA}
\par {\footnotesize $^{22}$ National Research University Higher School of Economics, Moscow, Russia;  Yandex School of Data Analysis, Moscow, Russia}
\par {\footnotesize $^{23}$ Fakult\"at f\"ur Physik, Ludwig-Maximilians-Universit\"at M\"unchen, M\"unchen, Germany}
\par {\footnotesize $^{24}$ University of Cincinnati, Cincinnati, OH, USA}
\par {\footnotesize $^{25}$ Physics Department, Brookhaven National Laboratory, Upton, NY, USA}
\par {\footnotesize $^{26}$ Laboratory for Nuclear Science, Massachusetts Institute of Technology, Cambridge, MA, USA}
\par {\footnotesize $^{27}$ Center for High Energy Physics, University of Oregon, Eugene, OR, USA}
\bigskip
\par {\footnotesize $^{a}$ Vladimir V. Gligorov acknowledges funding from the European Research Council (ERC) under the European Union's Horizon 2020 research and innovation programme under grant agreement No 724777 “RECEPT”}
\par {\footnotesize $^{b}$ Caterina Doglioni acknowledges funding from the European Research Council (ERC) under the European Union's Horizon 2020 research and innovation programme under grant agreement No 679305 “DARKJETS”}
\par {\footnotesize $^{c}$ Antonio Bovela acknowledges funding for US Department of Energy (Grant DE-SC0011726).}

\newpage
\setcounter{page}{1}

\section{Introduction and Scope}

Realizing the physics programs of the planned and/or upgraded high-energy physics (HEP) experiments over the next 10 years will require the HEP community to address a number of 
challenges in the area of software and computing. For this reason, the HEP software community has engaged in a planning process over the past two years, with the objective of 
identifying and prioritizing the research and development required to enable the next generation of HEP detectors to fulfill their full physics potential. The aim is to produce 
a Community White Paper (CWP) \cite{HSF2017} which will describe the community strategy and a roadmap for software and computing research and development in HEP for the 2020s. 
This activity is organised under the umbrella of the HEP software foundation (HSF). The LHC experiments and HSF have been specifically charged by the WLCG project, 
but have reached out to other HEP experiments around the world throughout the community process in order to make it as representative as possible.

The CWP process was carried out by working groups centered on specific topics. The topics of event reconstruction and software triggers are covered together in 
this document and have resulted from discussions within a single working group. The reconstruction of raw detector data and simulated data and its processing in 
real time represent a major component of today's computing requirements in high-energy physics. A recent projection \cite{Campana2016} of the ATLAS 2016 computing model 
results in $>85\%$ of the HL-LHC CPU resources being spent on the reconstruction of data or simulated events. This working group evaluated the most important components 
of next generation algorithms, data structures, and code development and management paradigms needed to cope with highly complex environments expected in high-energy 
physics detector operations in the next decade. New approaches to data processing were also considered, including the use of novel, or at least, novel to HEP, algorithms, 
and the movement of data analysis tasks into real-time environments. 

The remainder of this document is organized as follows. First we discuss how future changes including new and proposed facilities, detector designs, and evolutions in 
computing and software technologies change the requirements on software trigger and reconstruction applications. Second, we summarize current practices and identify the 
most limiting components in terms of both physics and computational performance. Finally we propose a research and development roadmap for the software trigger and event 
reconstruction areas including a survey of relevant on-going work for the topics identified in the roadmap. 

Of course any discussion of the computing challenges and priorities for software triggers and reconstruction necessarily overlaps with software domains covered by other 
CWP documents. Indeed, the critical role of real-time reconstruction in allowing the LHC data to be collected in the first place means that the requirements set out here 
will drive much of the R\&D across other areas, whether that be in the development of more performant math libraries, simplified but accurate detector descriptions, or 
new reconstruction algorithms based on machine learning paradigms. Such areas of overlap are noted wherever relevant in the text, and the reader is encouraged to refer to 
the other CWP documents for more details.

\section{Nomenclature}

This document will discuss software algorithms essential to the interpretation of raw detector data into analysis level objects in several contexts. 
Specifically, these algorithms can be categorized as: 
\begin{enumerate}
\item {\bf Online}: Algorithms, or sequences of algorithms, executed on events read out from the detector in near-real-time as part of the software trigger, typically 
on a computing facility located close to the detector itself.
\item {\bf Offline}: As distinguished from online, any algorithm or sequence of algorithms executed on the subset of events preselected by the trigger system, or generated 
by a Monte Carlo simulation application, typically in a distributed computing system.
\item
{\bf Reconstruction}: The transformation of raw detector information into higher level objects used in physics analysis. 
Depending on the experiment in question, these 
higher level objects might be charged particle trajectories (``tracks''), neutral or charged particle calorimeter clusters, Cherenkov rings, jets, and so on. 
 A defining characteristic of “reconstruction” which separates it from “analysis” is that the quality criteria used in the reconstruction to, for example, minimize the number of fake tracks, should be general enough to be used in the full range of physics studies required by the experimental physics program.
Reconstruction algorithms are also typically run as part of the processing carried out by centralized computing facilities.
\item
{\bf Trigger}: the online classification of events, performed with the objective of reducing either the number of events which are kept for further ``offline'' analysis, 
the size of such events, or both. In this working group we were only concerned with software triggers, whose defining characteristic is that they process data without a 
fixed latency. Software triggers are part of the real-time processing path and must make processing decisions quickly enough to keep up with the incoming data, possibly 
with the benefit of substantial disk buffers.
\item
{\bf Real-time analysis}: The typical goal of a physics analysis is to combine the products of the reconstruction algorithms (tracks, clusters, jets...) into complex 
objects (hadrons, gauge bosons, new physics candidates...) which can then be used to infer some fundamental properties of nature (CP asymmetry, lepton universality, Higgs couplings...). 
We define as real-time analysis any physics analysis step that goes beyond object reconstruction and is performed online within the trigger system, in certain cases 
using simplified algorithms to fit within the trigger system constraints. Real-time analysis techniques are so far quite experiment-specific. Techniques may include 
the selection and classification of all objects crucial to the calibration of the detector performance, evaluation of backgrounds, as well as physics searches that are 
otherwise impossible given limitations of data samples passing the trigger and saved for offline work.
\end{enumerate}

The online and offline algorithms have traditionally been viewed as related, but at least partly separate due to their differing goals and requirements. Because the online 
algorithms have to run on all events read out from the detector,\footnote{Of course online algorithms are a sequence, and some of the algorithms in this sequence may select 
events for processing by later algorithms. So not every online algorithm must run on every event read out from the detector, but when grouped together the ensemble of ``online'' 
algorithms process all events read out from the detector in near-real-time.}
they typically must be executed on dedicated computer facilities (e.g. a server farm) located 
near to the detector in order to avoid prohibitive networking and data transfer costs.\footnote{Note that this is not an argument about latency. A hardware fixed latency trigger 
has a set maximum time to evaluate any single event. A software trigger has no such constraint, and instead has some average processing time to evaluate events, given by a 
combination of the processing power available, the size of the disk buffers which can store data waiting to be analyzed, and the network bandwidth available to move events to this disk buffer. } 
Such dedicated farms typically have a small (if any) amount of disk space 
to buffer events while waiting for them to be processed, and the online algorithms must therefore be tuned to strict timing and memory consumption requirements. In contrast 
the offline algorithms run only on a subset of events which have been saved to long term storage. They must still execute within a reasonable time so that their output is made 
available for analysis in a timely fashion, and to fit the computing resources available to the experiment, but these pressures are generally much less severe than for online 
processing. In addition, online algorithms often run in dedicated frameworks, with additional layers of safety or control compared to their offline counterparts. 

Increasingly, however, the difficulties of maintaining these parallel software environments is driving online algorithms to become special cases of the offline ones configured 
for increased speed at the cost of precision, but otherwise relying on the same underlying codebase. This development is also driven by the desire to reduce systematic uncertainties 
introduced by having separate online and offline reconstruction, in particular for ``precision'' measurements.
This physics motivation to use offline algorithms online can lead to performance improvements in offline algorithms beyond what might have otherwise been achieved. In turn, such 
improvements free up offline resources, notably for producing the large samples of simulated events which will be needed in the HL-LHC period. We therefore assume that this trend 
will continue and intensify on the timescale considered in this document.

\section{New Challenges anticipated on the 5-10 year timescale}

This section summarizes the challenges identified by the working group for software trigger and event reconstruction techniques in the next decade. We have organized these 
challenges into those from new and upgrade accelerator facilities, from detector upgrades and new detector technologies, increases in anticipated event rates to be processed 
by algorithms (both online and offline), and from evolutions in software development practices.

\subsection{Challenges posed by Future Facilities }

Here we briefly describe some of the relevant accelerator facility upgrades and their impact on experimental data samples. These will be the basis of our discussion of how 
software trigger and event reconstruction algorithms must evolve.
\begin{enumerate}
\item 
LHC Run 3: Run 3 of the LHC is expected to last three years, starting in 2021, after the second long shutdown of the LHC. Having already exceeded the design luminosity of the 
LHC facility, there is no significant increase in instantaneous luminosity expected for the CMS and ATLAS experiments. The CMS and ATLAS experiments expect to accumulate up to 
300~fb$^{-1}$ of data each by the end of this run \cite{Bordry2016}. This is a nearly 10x increase over the samples of 13~TeV data collected through 2016. Both LHCb and ALICE will undergo 
major upgrades for Run 3 (described in the next section) : LHCb will have an instantaneous luminosity five times higher than in Run 2 \cite{LHCb2012}, while ALICE will 
upgrade \cite{ALICE2013} its readout and real-time data processing in order to enable the full 50~kHz Pb-Pb collision rate to be saved for offline analysis.
\item
High-luminosity LHC (HL-LHC): The HL-LHC project \cite{Zimmerman2009} is currently planned to begin operations in 2026. It is an upgrade to LHC, expected to result in an increase of up 
to a factor of 10 in instantaneous luminosity over the LHC design (so up to $10^{35}$~cm$^{-2}$s$^{-1}$). The beam energy of 14~TeV and 25~ns bunch spacing imply a considerable increase in the 
number of simultaneous collisions (pileup) seen by experiments. Operating scenarios under study are pileup of 140 or 200 for ATLAS \cite{ATLAS2015} and CMS \cite{CMS2015} at a 25~ns bunch 
spacing, both possibly with luminosity leveling techniques that would provide a relatively constant luminosity throughout a fill. In addition, LHCb \cite{LHCb2017} is planning a 
consolidation during LS3 for initial HL-LHC operations, with improvements to various detector components, followed by a potential later Upgrade II with a pileup of 50 to 60, to 
run from roughly 2030 onwards. The HL-LHC project is expected to run for at least 10 years. 
\item
Super KEK-B: The Super KEK-B facility \cite{Ohnishi2013}, together with the Belle-II experiment \cite{BelleII2010}, plans to achieve a 40 times increase in instantaneous luminosity over that 
achieved by the previous generation of $e^+e^-$ colliders operating on or near the Upsilon(4S) resonance (KEK-B and PEP-II). Super KEK-B should begin production data taking in 2018 and 
run until at least 2024 with a goal of 50~ab$^{-1}$ of data collected (e.g., nearly 60 billion $B$-$\bar B$ pair events recorded). 
\item
Long Baseline Neutrino Facility (LBNF): The LBNF project \cite{DUNE2016} is planning a high-intensity neutrino beamline from Fermilab to the SURF underground facility in South Dakota. 
Based on the NuMI beamline, LBNF plans a proton-beam power of 1.2~MW ($7.5\cdot 10^{13}$ protons per cycle) later upgraded to 2.4~MW ($1.5-2.0 \cdot 10^{14}$ protons per cycle). The facility expects 
to operate for 20 years starting 2025.
\item 
Linear Colliders (ILC and CLIC): two electron-positron collider projects are currently under study, the International Linear Collider (ILC), to be built in Japan and the Compact 
Linear Collider (CLIC) at CERN. The ILC will operate at a center of mass energy of 250-500~GeV. With a nominal luminosity of $1.47\cdot 10^{34}$~cm$^{-2}$s$^{-1}$ at 500~GeV and an expected raw data 
rate of 1~GB/s the two planned experiments will each accumulate up to 10~PB/year. Both colliders plan to run without any hardware trigger, requiring a fast and efficient prompt 
reconstruction and event building.
\item
Future Circular Collider (FCC): A 100 TeV facility is being studied as the next step in the energy frontier projects after HL-LHC \cite{Ball2014}. This could be realized as a 100 km 
circumference tunnel using 16 T magnets. Scenarios under discussion for such a facility include up to 1000 pileup events and the need to be hermetic up to a much larger eta 
(eg. $\eta=6$) than planned for the HL-LHC upgrades to the ATLAS or CMS experiments. 
\end{enumerate}
Several common themes are apparent from these facility plans. Accelerator operating conditions continue to evolve towards higher intensity and higher energy, as required to bring
new discovery potential. This means more complex and higher particle density environments from which signs of new physics must be extracted by trigger systems, event reconstruction 
algorithms and by analysis. This complexity brings new requirements and challenges to detector design as well as software algorithms, where detection efficiency needs to be maintained 
in more complex environments without increasing false-positive rates. 

For HL-LHC, the increased pileup of many interactions in a single crossing leads to several critical problems. The higher particle multiplicities and detector occupancies will lead 
to a significant slowdown in all reconstruction algorithms, from the tracking itself to the reconstruction in other devices such as the electromagnetic calorimeter and RICH detectors. 
In addition to making the algorithms slower, pileup also leads to a loss of physics performance, for example :
\begin{itemize}
\item Reduced reconstruction efficiency in Electromagnetic calorimeters;
\item Increased association of tracks to wrong primary vertices;
\item Reduced efficiency of identifying isolated electrons, muons, taus, and photons;
\item Reduced selection efficiencies for electrons and photons;
\item Reconstruction efficiencies for hadronic tau decays and b-jets;
\item Worse energy resolution for electrons, photons, taus, jets, and missing transverse energy;
\item Worse reconstruction of jet properties (substructure for top/$W$-tagging, quark/gluon discrimination, etc);
\end{itemize}

The central challenge for object reconstruction at HL-LHC is thus to maintain excellent efficiency and resolution in the face of high pileup values, especially at low object transverse momenta. 
Detector upgrades such as increases in channel density, high precision timing and improved detector geometric layouts are essential to overcome these problems. For software, particularly for 
triggering and event reconstruction algorithms, there is an additional need not to dramatically increase the event processing time. A comprehensive program of studies is required to assess the 
efficiencies and resolutions for various approaches in events with up to 200 pileup interactions. 

The increase in event complexity also brings a ``problem'' of overabundance of signal to the experiments, and specifically the software trigger algorithms. Traditional HEP triggers select a small 
subset of interesting events and store all information recorded in such events. This approach assumes first that only a very small fraction of collisions potentially contain interesting physics, 
second that the features of such events will be strikingly different than the features of ``uninteresting'' events, and third that discarding any information in an event would prevent later correcting 
any defects in the real-time processing, or preventing a full understanding of the process of interest. The evolution towards a genuine real-time processing of data has been driven by a breakdown 
in the first two assumptions and technological developments which means that the third assumption is no longer as worrying as it once was.

An illustrative example is the search for low-mass dark matter at the LHC, such as the  search for dark photons at LHCb \cite{Ilten2016}.  Since interactions between dark photons and Standard Model 
(SM) particles have very low cross sections, the probability of producing dark photons in proton-proton collisions is extremely small. Thus, discovering them at the LHC will require an immense 
number of proton-proton collisions and a highly efficient trigger. The key problem is that when the dark photon lifetime is small compared to the detector resolution, which is the case in much 
of the interesting parameter space, there is an overwhelming irreducible SM background from off-shell photons producing di-muon events. The current LHCb trigger configuration discards about 
90\% of potential dark photon decays, and a variety of other beyond the Standard Model and SM signals where the LHCb detector itself has good sensitivity. Once this stage is removed (and the 
luminosity is increased), the potential signal rate will increase by a factor of 50. However, the irreducible SM-background rate also increase significantly. Since offline storage resources 
are not expected to increase at nearly this rate, the data must be compressed much more than is now within the online system. In other words, techniques developed in the offline analysis for 
background must be integrated into the online software trigger. 

Similar considerations apply to low-mass dijet searches at ATLAS/CMS, where the enormous background rate from quantum chromodynamics (QCD), which grows linearly with pileup, limits the study of 
physics at the electroweak scale in hadronic final states given current techniques. Trigger output bandwidth limitations mean that only a tiny fraction of lower-energy jets can be recorded, and 
hence the potential statistical precision of searches involving these low energy jets is vastly reduced. This is again relevant in the context of dark matter searches, here for light mediators 
between quarks and dark matter particles with low coupling strength. Further details are described in \cite{aTLAs2017, CMS2016}.

These considerations also apply to other areas of the LHC physics program. For example, by Run~3 \cite{LHCb2014}, most LHC bunch crossings will produce charm hadrons at least partially in the LHCb 
acceptance, and all bunch crossings will produce strange hadrons detectable in the LHCb acceptance. Even more dramatically, in the HL-LHC era, a potential Upgrade II of LHCb will have to cope 
with multiple reconstructible charm hadron signals per bunch crossing, and even semileptonic beauty-hadron decays will become more abundant than can be stored offline. The ability of the LHC to continue 
improving our knowledge in these areas will therefore entirely depend on the ability to select the specific signal decay modes in question at the trigger level. Taken together, the overabundance 
of interesting signals, which mandates more complex event reconstruction at the trigger level, and the increasing event complexity, which makes the event reconstruction ever more expensive, will 
influence the design and requirements of trigger and reconstruction algorithms over the next decade.

\subsection{Challenges and Opportunities from Evolution in Experimental apparatus}

A number of new detector and hardware trigger concepts are proposed on the 5-10 year timescale in order to help in overcoming the challenges identified above. In many cases, these new technologies
bring novel requirements to software trigger and event reconstruction algorithms. These include:
\begin{enumerate}
\item
High-granularity calorimetry: Experiments including CMS (for HL-LHC), ILC, CLIC and FCC are proposing very high granularity calorimeters in order to better separate showers from closeby particles in a 
high-density (i.e., high-pileup) environment and thereby improve the jet energy resolution. This granularity brings significant computational challenges as there is much more information to process in 
order to fully take advantage of these devices. Efficient algorithms \cite{Marshall2015,Gray2016} are needed to fully optimize the ambiguity reduction capability of the signals from millions of channels 
within finite computing times. 

\item
Precision timing detectors \cite{Gray2017} for charged particles: Experiments including ATLAS, CMS and LHCb are pursuing timing detectors with precision approaching 30~ps. This information is another tool 
for dramatically reducing the effect of pileup on triggering and reconstruction algorithms in particular in the areas of tracking and vertex finding. Integrating timing information into Kalman filtering 
and vertex disambiguation algorithms can improve both physics performance and time-to-process events online and offline. Making current tracking plus new timing detectors effectively integrate into a 
performant 4D-detection system will necessitate the development of new reconstruction algorithms that make use of both spatial and timing information. 

\item
Hardware triggers based on tracking information: Hardware trigger systems designed to identify tracks down to 2 GeV of transverse momentum are a valuable tool for ATLAS and CMS \cite{ATLAS2015,CMS2015}. This 
technology will enhance the ability to trigger on a range of physics signatures including isolated leptons, multi-jet signatures and displaced vertices, as well as mitigating the effects of pileup 
on these objects. Once an event is triggered, the results obtained from these triggers can also be a valuable tool in the software trigger, for example by seeding and therefore speeding up the 
subsequent reconstruction. Similar systems, with an ability to go below 500 MeV of transverse momentum are also under consideration for the LS3 Consolidation and Upgrade II of LHCb, and could be 
particularly valuable in the study of strange hadrons if the thresholds could be reduced down to 100 MeV of transverse momentum.

\item
Data streaming techniques: Experiments with no hardware trigger allow a software trigger to see all data before events are rejected from further processing. There is a clear advantage in physics 
capability if the data streaming capability is sufficient and if software triggers are efficient, effective and reliable enough for the task. There is always an advantage if additional algorithms 
can be run on an event before a decision must be taken about its importance. In the case of LHCb this means a facility and algorithms that are capable of processing with a sustained throughput 
of 30~MHz \cite{LHCb2014}. Similarly, the Alice experiment plans a 50~kHz interaction rate with no, or simple minimum bias, hardware trigger, and will stream 3~TB/s from the TPC in a common 
online-offline computing system \cite{ALICE2015}.
\end{enumerate}

\subsection{Challenges from Event rates and real-time processing}

Trigger systems for next-generation experiments are evolving to be more capable, both in their ability to select a wider range of events of interest for the physics program of their experiment, 
and their ability to stream a larger rate of events for further processing.  ATLAS and CMS both target systems where the output of the hardware trigger system is increased by 10x over the current 
capability, up to 1~MHz \cite{ATLAS2015,CMS2015}. In other cases, such as LHCb, the full collision rate (between 30 to 40~MHz for typical LHC operations) will be streamed to real-time or quasi-realtime 
software trigger systems. It is interesting to note that because the ATLAS/CMS events are O(10) times larger than those of LHCb (roughly O(1) vs O(0.1) MB/event), the resulting data rates are 
rather similar, namely 1-5~TB/s. 

This enhanced capability naturally increases the demands on software trigger algorithms and offline reconstruction algorithms. In many cases, the current trigger bandwidth of experiments is limited 
by the offline processing and storage capabilities of an experiment rather than its ability to physically write out data to disk or tape for further processing.  This can be due either to the time 
to fully reconstruct events for analysis (corresponding to a fixed set of CPU resources)  or by the size of the analysis data itself (corresponding to a fixed set of disk resources). Current 
experiments are therefore constantly working to reduce the CPU needs to reconstruct events and storage needs for analyzing them, through a combination of code improvements (refactoring, vectorization, 
exploitation of optimized instruction sets) or through entirely rewriting algorithms. 

This is an ongoing process that continues to yield throughput improvements, aided by improved code analysis tools, modern compilers and other tools. For many experiments, there is also a potential 
tradeoff between physics quality and CPU needs. A typical example is that track reconstruction requires less CPU if the transverse momentum threshold for tracks to be identified is raised, thereby reducing the 
overall combinatorics. This sort of CPU improvement is almost never desirable as it reduces the overall physics output. Instead software trigger and reconstruction applications are pressed to 
include more and more algorithms over time. Typical examples for ATLAS and CMS are new jet-finding, isolation, or pileup subtraction approaches, and, more generally, algorithms developed targeting 
specific physics use cases, for example in the trigger-level search for low-mass hadronic resonances in ATLAS described in \cite{Abreu2014}. Conversely, the need to reconstruct higher multiplicity, or 
increasingly soft, decays challenges LHCb’s applications.

Recent examples of significant storage reductions for the analysis data tier include the CMS MiniAOD \cite{Petrucciani2015} and the ATLAS xAOD \cite{Eifert2015}, both deployed for LHC Run 2 analysis. 
Improvements can be achieved by refinements in physics object selection (e.g., saving less uninteresting information), and by enhanced compression techniques (using lossless or lossy methods). 

Real-time analysis is the only solution for those signals which are so abundant that they cannot all be saved to disk, or for which discriminating against backgrounds requires the best possible 
detector calibration, alignment, reconstruction, and analysis. In this case, some or all of the output of the software trigger system also serves as the final analysis format. Its development is 
justified by the need to conduct the broadest possible program of physics measurements with our existing detectors. This is critical for two reasons: first, because we do not want to miss any 
signatures of New Physics around the electroweak scale, whether direct or indirect, but also because, even if the New Physics scale lies beyond the reach of current detectors, we must probe the 
widest possible parameter space in order to motivate and guide the design of future colliders and experiments. This is particularly true given the cost and long timescale of such future facilities.

An early version of this approach consisted of keeping only a limited set of physics objects (e.g., jets) as computed in real-time processing, and proof-of-concept implementations exist since LHC 
Run~1 \cite{Aaij2016,Abreu2014,CMS2016}. In order to perform precision measurements in real-time, however, it is critical to be able to keep data long enough (hours or days, depending on the experiment) 
to perform quasi-real-time calibrations and a final offline analysis quality reconstruction in the trigger system itself. This approach was commissioned by LHCb in 2015. As a result, roughly one 
third of the LHCb experiment’s Run~2 trigger selections now use the real-time reconstruction to reduce the amount of data kept for further analysis. Advantages of these approaches include an order 
of magnitude reduction in data volume from not saving the raw detector data, potentially reduced systematic errors in analysis due to differences in algorithms or calibrations used in the online 
and offline processing, and a reduction or elimination of the need for offline processing. 


These advantages allow an entire category of physics to be probed by HL-LHC experiments that would not otherwise be considered. Among the challenges posed by these approaches are the need for 
very robust event reconstruction algorithms, the need to derive detector calibrations sufficient for final analysis within this short processing window, and the need to plan analyses sufficiently 
in advance of data taking so that the choices made in the real-time analysis will be robust against eventual systematics studies.

\subsection{Challenges from Evolutions in Computing technology}

This section summarizes recent, and expected, evolutions in computing technologies. These are both opportunities to move beyond commodity x86 technologies, which HEP has used very effectively over 
the past 20 years, and significant challenges to continue to derive sufficient event processing throughput per cost to enable our physics programs at reasonable computing cost.  A full description 
of this technology evolution and its effect on HEP is beyond the scope of this document \cite{Bird2014}. Here we identify the main technology changes identified as driving out research and development 
in the area of software trigger and event reconstruction:
\begin{itemize}
\item
{\bf Increase of SIMD capabilities}: The size of vector units on modern commodity processors are increasing rapidly. While not all algorithms can easily be adapted to benefit from this capability, 
large gains are possible where algorithms can be vectorized. Essentially all HEP codes need modifications, or large scale refactoring, to effectively utilize SIMD capabilities.
\item
{\bf Evolution towards multi- or many-core architectures}: The current trend is to move away from ever faster processing cores towards more power efficient and more numerous processing cores. This 
change has already broken the traditional ``one-core-one-event'' model in many HEP frameworks and algorithms. As core counts increase, a larger number of algorithm developers, instead of just those 
developing the most resource intensive algorithms, will need to incorporate parallelism techniques into their algorithm implementations.   
\item
{\bf Slow increase in memory bandwidth}: Software trigger and event reconstruction applications in HEP are very memory intensive, and I/O access to memory in commodity hardware has not kept up with 
CPU capabilities. To evolve towards modern architectures, including the effective use of hierarchical memory structures, HEP algorithms will need to be refactored or rewritten to considerably reduce 
the required memory per processing core. 
\item
{\bf Rise of heterogeneous hardware}: Evolution in HEP algorithms has long taken advantage of a single dominant commodity computing platform (x86 running Linux).  More and more studies have shown 
or are investigating the throughput benefits of using low-power systems, GPUs, FPGA systems for critical pieces of processing. Software trigger and event processing algorithms are those that may 
benefit the most in HEP from the effective use of these technologies if they are able to adapt to the requirements of using these systems effectively.
\item
{\bf Possible evolution in facilities}: HEP facilities are likely to evolve both due to the increased architectural variability of affordable hardware and due to evolution in data science techniques. 
One example of the latter is an analysis center whose design is driven by data science techniques and technologies. These technologies (e.g., Hadoop, Spark) are under investigation in HEP for analysis 
and may change the way HEP data centers are resourced. A particular example of how this will impact trigger or reconstruction algorithms is that of physics object identification algorithms. These 
are frequently rerun by analysts in order to include the most recent version developed by the collaboration in their analysis.
\end{itemize}
Evolution in computing technology is generally a slow but continual process. However, architectures available today can provide necessary development platforms for trigger and event reconstruction 
developers to adapt codes to be better suited to future architectures.

\subsection{Challenges from Evolutions in Software technology}

The status and evolution of software development in HEP is the subject of the Software Development CWP working group. In this section, we briefly discuss some of the issues and opportunities of 
particular relevance to software trigger and event reconstruction work.

The move towards open source software development and continuous integration systems brings a number of important opportunities to assist developers of software trigger and event reconstruction 
algorithms. Continuous integration systems have already brought the ability to automate code quality and performance checks, both for algorithm developers and code integration teams. Scaling 
these up to allow for sufficiently high statistics checks is among the still outstanding challenges. While it is straightforward to test changes where no regression is expected, fully developed 
infrastructure for assisting developers in confirming the physics and technical performance of their algorithms during development is a work in progress.

As the timescale for experimental data taking and analysis increases, the issues of legacy code support increase. In particular, it seems unaffordable to imagine rewriting all of the software 
developed by the LHC experiments during the long shutdown preceding HL-LHC operations. Thus, as the HL-LHC run progresses, much of the code base for software trigger and event reconstruction 
algorithms will be 15-30 years in age. This implies an increased need for sustainable software development and investment in software education for experimental teams.

Code quality demands increase as traditional offline analysis components migrate into trigger systems, or more generically into algorithms that can only be run once. As described above, 
this may be due to either the prohibitive cost of rerunning algorithms over large data sets, or due to not having retained sufficient data to rerun algorithms (e.g., not retaining the full 
raw data). Algorithms in the software trigger and event reconstruction areas are very frequently contributed to by a large community of physicists rather than expert programmers. In many 
cases, the most sensitive algorithms may be checked and optimized by programming experts. This has so far satisfied the need of reducing the total computing resources needed by experiments, 
as typically only a few algorithmic components dominate the overall computing need. However, this approach would be currently impossible to carry out across the entire reconstruction code 
stack. As the complexity (and number) of real-time algorithms increases, the need for training as well as code validation and regression checking increases considerably. 

These challenges are further complicated by a growing diversity of developers in the software trigger and event reconstruction areas. There is a generally growing gap between the basic 
programming techniques which students learn in their undergraduate courses and the state-of-the-art programming techniques required to fully exploit the power of emerging parallel hardware 
architectures. Software development methods and programming techniques evolve particularly quickly and often developers are self taught on current techniques. The experiment software environment 
must facilitate contributions from a range of developers while adjusting to challenges of a more and more complex online and offline software toolkit.

\section{Current Approaches}

In this section, we summarize current practices and resource requirements for software trigger and event reconstruction implementation and infrastructure needs. These include the scale of resource 
requirements for these activities, the type of data structures and data contents kept for custodial storage and analysis use, and calibration technique requirements across different experiments in 
high-energy physics. For each topic, we will briefly introduce the issues and then summarize experiment specific implementation details where available. This is meant to provide a representative 
range of approaches and to illustrate where the most challenging aspects are. Approaches from all experiments are not always included, as this information is summarized 
from working-group contributions. 

\subsection{Computing Resource Requirements}

The table below summarizes the online and offline computing requirements of current and future experiments. Not all quantities are relevant for all experiments, and many items evolve depending 
on where in the lifecycle an experiment is. For example, disk requirements depend both on the number of events collected in a given running period but also on how long the experiment has been 
collecting data still of relevance for current analysis activities. More information about each online and offline computing system can be found in the 
references \cite{ALICE2015, Allton2017, ATLAS2015, CMS2015, DUNE2015, DUNE2017, LHCb2017, Miyamoto2015, Richter2016}.

The columns are defined as follows:
\begin{enumerate}
\item Online CPUs: Approximate size (or fair-share in case of a shared resource) of the real-time processing facility available for use by the experiment.
\item Offline CPUs: Approximate size (or fair-share in case of a shared resource) of the offline processing facility available for use by the experiment. For LHC experiments we have given the 
2017 pledged CPU amount including both Tier0 and distributed Tier1 and Tier2 resources. These resources are used for Monte Carlo simulation, analysis and event reconstruction processing. 
Thus, these numbers only provide an indication of the CPU requirements for event reconstruction.
\item Input rate to software trigger: Approximate rate of events seen by software trigger. Typically this is the output rate of the hardware trigger.
\item Event rate for analysis: Approximate rate of events saved by the experiment for custodial storage. Typically this is the rate of events saved by the software trigger system.
\item RAW data size: Approximate size of typical physics events saved for offline processing. While the analysis tier size for an experiment is only loosely connected to the raw data size, 
this measure does provide a measure of the total data volume for the offline event processing. As discussed above, there are also frequently reduced data formats saved by trigger systems 
instead of the full raw data. These are typically much smaller than the sizes provided here.
\item Analysis data size: Approximate size of typical physics events in the format used most frequently for analysis (e.g., xAOD in ATLAS, miniAOD in CMS).
\item Offline disk: The total disk available for production and analysis activities. This is expected to scale primarily with the total size of the analysis data.
\end{enumerate}

\afterpage{
\clearpage
\begin{landscape}
\begin{table}
\begin{tabular}{|l|p{20mm}|p{20mm}|p{20mm}|p{20mm}|p{20mm}|p{20mm}|p{20mm}|} \hline
Experiment &
Online CPUs (kHS06) /GPUs  &
Offline CPUs (kHS06) &
Input rate to software trigger (kHz) &
Event rate for analysis (kHz or GB/s) &
RAW data size (MB/evt) &
Analysis data size (MB/evt) &
Offline disk (PB) \\ \hline
CMS (2017) &
500&
1729&
100&
1&
1.5&
~0.03&
123 \\  \hline
CMS (HL-LHC) &
&
&
750&
5-7.5&
4.2-4.6&
&
\\ \hline
ATLAS (2017) &
28k \newline CPU cores&
2194&
100&
1&
1&
&
172\\ \hline
ATLAS (HL-LHC) &
&
&
1000&
10&
5&
&
\\ \hline
LHCb (2017)&
25k \newline CPU cores&
413&
1000&
0.7 Gb/s&
0.07&
&
35\\ \hline
LHCb (Run 3-4) &
&
&
30000&
$>2$ GB/s&
0.13&
&
\\ \hline
ALICE (2017) &
9k \newline CPU cores, \newline 180 GPUs &
805&
0.3\newline (central \newline Pb-Pb)&
&
100\newline (central \newline Pb-Pb) &
&
66 \\ \hline
ALICE (Run 3) &
&
&
50&
20 GB/s&
60&
&
60/yr \\ \hline
Belle II (2021) &
6400 \newline CPU cores&
600&
30 kHz&
10 kHz&
0.1&
0.01&
60\\ \hline
ProtoDUNE&
&
5&
&
0.6 GB/s&
60&
&
\\ \hline
DUNE (0 supp)&
&
&
&
11 MHz&
$1.5e^{-4}$&
&
54/yr \\ \hline
ILC (ILD,SiD)&
250&
400&
1 GB/s&
1 GB/s&
1&
0.01&
150 \\ \hline
\end{tabular}
\caption{The LHCb resource requirements for Run 3 and 4 are currently under review, and will be agreed following publication of a computing model document later in 2018. 
The figures given here are minimal requirements below which physics performance is already known to degrade in an unacceptable manner.}
\end{table}
\end{landscape}
\clearpage
}

This table illustrates that there is substantial commonality in the scale of computing needed for current generation of experiments. It is also clear that there is a large increase 
in the scale of data expected for the next generation of experiments. This is true both in terms of event rate and event size.

\subsection{Analysis Data tiers and data structures}
Here we summarize data structures either consumed by, or produced by, the software trigger and event reconstruction applications in some current HEP experiments. These are meant to be 
representative of current practices and not inclusive of all approaches used in HEP. Historically, these are the raw data format and analysis data format, respectively. Recent work has 
led to some evolution in what sort of data structures are used in some experiments. We summarize the approach taken and issues observed by different experiments.

\vskip 0.5cm
\noindent
{\bf LHCb}: Major advances were made in the LHCb trigger system in Run 2. Increasing the number of (logical) cores used in the trigger farm to 50000, deploying faster reconstruction algorithms, 
and data-caching (described below) now allow the multilayer software trigger system to execute the offline reconstruction algorithms. LHCb now splits its 0.7~GB/s of data written to permanent 
storage into two distinct streams. The full stream persists all low-level data as in Run 1. The turbo stream persists a user-defined selection of high-level information, typically a small 
subset of reconstructed objects in each selected event. This makes it possible to keep a much larger number of events, with the cost being that the lower-level information is lost. 
Several LHCb publications have now demonstrated that this reduced-file-size approach works in practice, with typical event size reduction being one order of magnitude with respect to the raw event. 

\vskip 0.5cm
\noindent
{\bf CMS}: The data formats used by CMS for event data are built from custom written C++ classes organized into a ROOT TTree-based structure by the CMS framework. The RAW data is largely 
composed of a packed byte format organized and segmented according to the detector readout electronics layout. Analysis data are made up of more complex objects, used also in the 
reconstruction and analysis algorithms themselves. Particle flow constituent objects and high level physics objects (e.g. muon, electrons, jets) are made of classes derived from a 
common base class in order to make combination of objects in a straightforward and efficient manner.

Formats in Run~2 have evolved to include more compact data structures (e.g., the miniAOD formats) and for a few trigger streams the concept of ``scouting''. Instead of keeping raw detector data, 
the scouting data structures keep only a small set of high-level objects as derived in the CMS software trigger. This makes the data format very compact (about 1\% of a full raw data event) 
and allows for a very high rate of data to be kept. CMS has also used data structures with lossy-compression algorithms (e.g., saving variables using fewer than 32 bits) to reduce the size of 
its ``MIniAOD'' data tier, now widely used in analysis. This type of approach is widely applicable where the detector configuration or reconstruction algorithm itself limits resolution 
to be far worse than that implied by the usual 32-bit or 64-bit data types. While the custom structures developed by each experiment so far are unlikely to be completely generalizable, 
we would still benefit from a field-standard, or data-science standard library containing those parts which are in common.  CMS is also investigating how less complex data structures 
can allow for even small data formats that match the needs of the final analysis stages.

\vskip 0.5cm
\noindent
{\bf ATLAS}:In Run 2, a reduced analysis object stream has been added, containing only high-level trigger jet objects, to be 
used for the search of low-mass hadronic resonances in trigger-level analyses. This stream requires roughly 1\% of the full HLT bandwidth, allowing for data taking rates comparable to the 
full offline physics stream \cite{ATLASTwiki} 

\subsection{Most resource consuming algorithm}
{\bf ALICE}: Reconstruction and compression of data from the ALICE TPC is the dominant component of the ALICE event reconstruction and the driving feature behind the design of the O2 
facility \cite{ALICE2015} for event reconstruction starting in LHC Run 3. The reconstruction of particle trajectories from the ALICE TPC is done by cellular automation and Kalman filter algorithms. 
In the Run 2 high-level trigger, these algorithms run on GPUs as part of a pipelined system.

\vskip 0.5cm
\noindent
{\bf LHCb}: The specific forward geometry of LHCb, with a dipole magnet, and its use of RICH detectors for particle identification, have a significant impact on both the time cost of 
reconstruction algorithms and their utility in triggering and real-time analysis. The largest components are pattern recognition and Kalman fitting of tracks, which run on all events, and 
calorimeter and RICH reconstruction, which run on a subset of events. The overall CPU cost for these algorithms is roughly the same, considering the different fractions of events they run on. 
The full tracking cost greatly depends on the transverse momentum range of tracks being searched for. The RICH reconstruction is expensive, but because many decays of interest include kaons, 
its results reduce the time spent in later processing for combining tracks into displaced vertices.

\vskip 0.5cm
\noindent
{\bf CMS}: The computing resource needs of trigger and reconstruction algorithms is a strong function of event complexity (or pileup) in CMS. In Run 2 conditions, the track pattern recognition 
is the largest single CPU contributor in CMS reconstruction. Kalman fitting \cite{CMS2014}, particle flow algorithms \cite{CMS2017} and HCAL local reconstruction algorithms are also a considerable 
fraction of the total offline reconstruction CPU budget. The importance of these algorithms is also true for online trigger algorithm configurations where a simplified particle flow technique
 is used in the final event selections. As they are carried out in each event online, the calorimetric raw data unpackers and algorithms for pileup rejection at the detector level (e.g. to form 
reconstructed detector clusters) are also quite important online. Both online and offline, CMS has implemented a low track transverse momentum threshold (200-300 MeV) in order to achieve the best possible particle 
flow performance. This has a large impact on the CPU needs of CMS, not only in tracking, but elsewhere in the reconstruction where the algorithm complexity scales with the number of tracks 
reconstructed. For HL-LHC, CMS has the additional computational challenge of clustering and particle flow with its high-granularity calorimeter.

\subsection{Calibration techniques and requirements}
{\bf LHCb}: the full detector alignment and calibration of all sub-detectors must continue to be performed continuously in real-time, and the real-time monitoring framework must also be maintained 
in order to enable problems to be quickly identified and fixed. The calibrations are performed \cite{LHCbRTCalib2016} at a certain frequency:
\begin{itemize}
\item
A few times per year: Muon system alignment, RICH mirror alignment, Fine calorimeter calibration 
\item
Start of each LHC fill: Vertex detector alignment, Full tracker alignment, Straw-tube tracker gas calibration, Coarse calorimeter calibration
\item
Multiple times per fill: RICH refractive index calibration
\end{itemize}
In order to perform the calibrations in real-time, LHCb exploits the trigger farm to run the calibration tasks in parallel with trigger tasks. This enables e.g. the full tracker alignment to be 
completed in around 8 minutes. The calibration jobs are automated and return an updated set of constants which are compared to the result of the previous calibration; if the change goes beyond a 
certain threshold the constants are updated and an expert is alerted to validate the update. The jobs themselves are built on the same codebase as the rest of the LHCb reconstruction, and the 
monitoring uses the same setup as the general detector monitoring.

\vskip 0.5cm
\noindent
{\bf CMS}: The prompt calibration loop is an automated process, runs as part of the CMS Tier-0 facility using a small subset of the total data volume saved by the CMS trigger. Example calibrations 
include the assessment of bad detector channels for each run as well as the global tracker alignment. Results are typically available within 24 hours and thus are used in the full prompt reconstruction 
processing of the CMS data. Other calibrations, including the calorimeter light output calibrations, are performed outside of the Tier-0 infrastructure but in time for the prompt reconstruction 
processing on a regular interval. Finally, calibrations requiring higher statistics are performed either as the LHC data taking period progresses (e.g. performed once adequate statistics are obtained) 
or after the end of the run to achieve the ultimate detector performance required by analysis (e.g, tracker module level alignment and full energy scale calibration of the electromagnetic calorimeter). 

\vskip 0.5cm
\noindent
{\bf ATLAS}: For real-time analyses with trigger jets, reconstruction and calibration are  kept as close as possible to the full reconstruction. 
Dedicated calibrations mitigating the effect of pile-up and restoring the hadronic energy scale are applied to these partially recorded events, making the properties of jets reconstructed at 
the HLT comparable to those of jets reconstructed from full events. These calibrations account for differences between fully recorded jets as well as for missing information from detectors 
other than the calorimeters. In the case of jets during the first part of Run 2, only calorimeter information is available to the HLT for jet reconstruction and calibration. It is foreseen 
that tracking information can be added already during Run 2 using the Fast TracKer (FTK). The jet reconstruction procedure for trigger jets is summarized in \cite{Abreu2014}. 

\vskip 0.5cm
\noindent
{\bf ALICE}: Real-time calibration and data quality monitoring are critical parts of the O2 system for Run 3 as the data volume processed is too large to be redone in a later, truely offline, processing.

\section{Research and Development Roadmap}

This section describes the proposed research and development roadmap defined in the working group. We identify seven broad areas to be critical for software trigger and event reconstruction 
work over the next decade. These  are:
\begin{itemize}
\item Enhanced vectorization programming techniques
\item Algorithms and data structures to efficiently exploit many-core architectures
\item Algorithms and data structures for non-x86 computing architectures (e.g., GPUs, FPGAs)
\item Enhanced quality assurance (QA) and quality control (QC) for reconstruction techniques
\item Real-time analysis
\item Precision physics-object reconstruction, identification and measurement techniques
\item Fast software trigger and reconstruction algorithms for high-density environments
\end{itemize}
For each area, we identify the overall goals of the research, as well as short, medium and long term milestones. These can be viewed as goals that should be achieved in advance of next-generation 
experiments. The short and medium term milestones are intended to be achievable on a timescale to inform software, computing and trigger technical design reports where possible.

\subsection{Roadmap area 1: Enhanced vectorization programming techniques}
\noindent {\bf Motivation}: HEP developed toolkits and algorithms typically make poor use of vector units on commodity computing systems. Improving this will bring speedups to applications running 
on both current computing systems and most future architectures.

\noindent {\bf Overall goal}: To evolve current toolkit and algorithm implementations, and best programming techniques to better use SIMD capabilities of current and future computing architectures.

\vskip 0.5cm
\noindent {\bf Short-term goals}: Identify best practices and documented examples of how HEP code was improved to increase vectorization performance via series of developer meetings. Adopt and apply 
industry tools for code performance analysis to identify tools of particular importance for investigation.

\vskip 0.5cm
\noindent {\bf Medium-term goals}: Initiate work to make measurable improvement to vectorization performance of HEP code stacks via software toolkit improvements or rewrites. Continue developer 
discussions while toolkits are improved to facilitate co-development where common approaches to improving different toolkits are possible. 

\vskip 0.5cm
\noindent {\bf Long-term goals}: Demonstrate and facilitate widespread experiment adoption of new or improved toolkits.

\subsection{Roadmap area 2: Algorithms and data structures to efficiently exploit many-core architectures}
\noindent {\bf Motivation}: Computing platforms are generally evolving towards having more cores in order to increase processing capability. This evolution has resulted in multi-threaded frameworks 
in use, or in development, across HEP. Algorithm developers can further improve throughput by being thread safe and enabling the use of fine-grained parallelism. 

\vskip 0.5cm
\noindent {\bf Overall goal}: To evolve current event models, toolkits and algorithm implementations, and best programming techniques to improve the throughput of multithreaded software trigger and 
event reconstruction applications.

\vskip 0.5cm
\noindent {\bf Short-term goals}: Identify the key lessons of the work done to make current toolkits thread-safe or at least ``thread friendly''. Understand and document what conceptual limitations
a thread-safe framework imposes on reconstruction and selection logic, and what kind of workflow scheduling it requires. Identify what classes of reconstruction and selection algorithms are able 
to efficiently work in a multi-threaded framework and which are not. 

\vskip 0.5cm
\noindent {\bf Medium-term goals}: Document best practice for thread safe computing in a way which allows new collaboration members to efficiently develop in such a framework. Develop a toolkit 
to logically express proposed new algorithms in terms of data sources, sinks, and consumers, which can automatically analyze an algorithm and establish how thread-safe its logic is.

\vskip 0.5cm
\noindent {\bf Long-term goals}: Introduce training on multi-threaded and more generally parallel algorithm development in physics syllabi, building on the experience gained and the toolkits developed earlier.

\subsection{Roadmap area 3: Algorithms and data structures for non-x86 computing architectures (e.g., GPUs, FPGAs)}
\noindent {\bf Motivation}: Computing architectures using technologies beyond CPUs offer an interesting alternative for increasing throughput of the most time consuming trigger or reconstruction 
algorithms. Such architectures (e.g., GPUs, FPGAs) could be easily integrated into dedicated trigger or specialized reconstruction processing facilities (e.g., online computing farms).

\vskip 0.5cm
\noindent {\bf Overall goal}: To demonstrate how the throughput of toolkits or algorithms can be improved through the use of new computing architectures in a production environment. 

\vskip 0.5cm
\noindent {\bf Short-term goals}: Develop reliable and portable benchmarking for mixed architectures which properly accounts for I/O overheads between them. Using this, identify, and develop 
prototypes for, event reconstruction and software trigger algorithms where specialized hardware is likely to bring significant improvement in metrics such as total event throughput per cost or throughput per Watt. 

\vskip 0.5cm
\noindent {\bf Medium-term goals}: Demonstrate programming models and software toolkits appropriate for a heterogeneous computing environment. Considerations include facilitating high-level 
code-reuse between architectures, construction of appropriate data structures, and adoption of externally developed toolkits, such as mathematical libraries, that provide significant performance 
improvements on certain architectures. Deploy prototypes for limited scale operational tests where resources are available in a controlled fashion (e.g., experiment trigger computing facilities).

\vskip 0.5cm
\noindent {\bf Long-term goals}: Demonstrate ability of event reconstruction applications to reliably use and benefit from heterogenous computing facilities on a distributed computing system. 
Define cost-benefit metrics to use to guide facility providers for providing the largest event throughput in cases where HEP controls the mix of hardware to be purchased.

\subsection{Roadmap area 4: Enhanced QA/QC for reconstruction techniques}
\noindent {\bf Motivation}: HEP experiments have extensive continuous integration systems, including varying types of code regression checks. These are typically maintained by individual experiments 
and have not yet reached the scale where statistical regression checks, as well as technical and physics performance can be enabled for each proposed software change.

\vskip 0.5cm
\noindent {\bf Overall goal}: Enable the development, automation, and deployment of extended QA and QC tools and facilities for software trigger and event reconstruction algorithms. 

\vskip 0.5cm
\noindent {\bf Short-term goals}: Discuss integration and testing systems currently used by experiments to formulate requirements for scope and scale needed for a common integration system 
capable of providing developers feedback on trigger and event reconstruction outputs according to physics and technical metrics.

\vskip 0.5cm
\noindent {\bf Medium-term goals}: Develop and demonstrate a scalable system for use by multiple experiments based on industry standard continuous integration tools. Define requirements for 
regression and validation techniques (for Monte Carlo and data studies) given evolution towards heterogeneous hardware and real-time calibrations.

\vskip 0.5cm
\noindent {\bf Long-term goals}: Develop and demonstrate next-generation tools needed for regression testing, software integration and validation and data quality related tasks. 

\subsection{Roadmap area 5: Real-time analysis}

\noindent {\bf Motivation}: Real-time analysis techniques are being adopted to enable a wider range of physics signals to be saved by the trigger for final analysis. As rates increase, these 
techniques can become more important and widespread by enabling only techniques such as saving only reconstructed event information or only
the parts of an event associated with the signal candidates, reducing the required disk space substantially.

\vskip 0.5cm
\noindent {\bf Overall goal} : Evaluate and demonstrate the tools needed to facilitate real-time analysis techniques. Research topics include compression and custom data formats; toolkits for real-time 
detector calibration and validation which will enable full offline analysis chains to be ported into real-time; and frameworks which will enable non-expert offline analysts to design and deploy 
real-time analyses without compromising data taking quality. 

\vskip 0.5cm
\noindent {\bf Short-term goals}: Discuss the ongoing real-time analysis frameworks being built by different HEP experiments and establish areas of commonality. Understand the extent to which 
cross-experiment toolkits can help with these areas. 

\vskip 0.5cm
\noindent {\bf Medium-term goals}: Develop common toolkits for enabling real-time analysis across experiments, drawing on experience and collaboration with real-time applications in industry 
wherever possible. Develop a framework which enables non-experts to design and deploy real-time analyses without threatening the integrity of data taking.

\vskip 0.5cm
\noindent {\bf Long-term goals}: Begin to include real-time analysis requirements in the design experiments (in particular high luminosity hadron colliders such as FCC). This means explicitly 
optimizing detector hardware not for physics which can be done with events we can afford to store to disk, but optimizing for physics which can be done in real-time.

\subsection{Roadmap area 6: Precision physics-object reconstruction, identification and measurement techniques}

\noindent {\bf Motivation}: The central challenge for object reconstruction at HL-LHC is thus to maintain excellent efficiency and resolution in the face of high pileup values, especially at 
low object transverse momentum. Both trigger and reconstruction approaches need to exploit new techniques and higher granularity detectors to maintain or even improve physics measurements in the future.
Reconstruction in very high pileup environments, such as the HL-LHC or FCC-hh, may also greatly benefit from adding timing information to our detectors, in order to exploit the finite beam crossing time during which interactions are produced.

\vskip 0.5cm
\noindent {\bf Overall goal}: Develop and demonstrate tools needed to efficient techniques for physics object reconstruction and identification in complex environments.

\vskip 0.5cm
\noindent {\bf Short-term goals}: Identify areas where either new toolkits based on either novel techniques or new detector designs, are likely to achieve significant physics quality 
improvements especially in very dense (e.g., high pileup) environments at facilities including HL-LHC and FCC-hh. Known candidates are charged tracking techniques including precision timing 
detector information, jet imaging techniques, and particle-flow algorithms that exploit high-precision calorimetry. 
 
\vskip 0.5cm
\noindent {\bf Medium-term goals}: Development and demonstration of algorithms and integrated software packages that are efficient and performant in complex environments on planned detector 
configurations for HL-LHC. Understand interplay between new information and traditional observables to determine how physics measurables should be best derived. Development of algorithms to 
optimize splitting events containing different sets of objects to obtain the best balance between data storage overhead and data processing overhead during physics analyses.

\vskip 0.5cm
\noindent {\bf Long-term goals}: Deploy algorithms in experimental software stacks as they mature.

\subsection{Roadmap area 7: Fast software trigger and reconstruction algorithms for high-density environments}

\noindent {\bf Motivation}: Future experimental facilities will bring a large increase in event complexity. The scaling of current-generation algorithms with this complexity must be improved 
to avoid a large increase in resource needs. In addition, it may be desirable or indeed necessary to deploy new algorithms, including advanced machine learning techniques developed in other fields, 
in order to solve these problems.

\vskip 0.5cm
\noindent {\bf Overall goal}: Evolve or rewrite existing toolkits and algorithms focused on their physics and technical performance at high event complexity (e.g. high pileup at HL-LHC). Most 
important targets are those which limit expected throughput performance at future facilities (e.g., charged-particle tracking). A number of such efforts are already in progress across the community.

\vskip 0.5cm
\noindent {\bf Short-term goals}: Identify additional areas where substantial gains in event reconstruction and software trigger algorithms may be obtained by either a large-scale reimplementation 
of existing algorithms or by the use of a new algorithmic approach (including machine learning concepts). Possible areas of investigation include improved memory locality for algorithms and data structures.

\vskip 0.5cm
\noindent {\bf Medium-term goals}: Develop and demonstrate new toolkits. Evaluate their effectiveness against current approaches using both physics driven and event throughput per computing cost metrics. 
\vskip 0.5cm
\noindent {\bf Long-term goals}: Deploy algorithms in experimental software stacks as they mature. It is particularly important to test new approaches using data driven studies to demonstrate 
robustness against changing detector and accelerator operating conditions.

\section{Conclusions}

The next decade will see the volume and complexity of data being processed by HEP experiments increase by at least one order of magnitude. While much of this increase is driven by the planned 
upgrades to the four major LHC detectors, new experiments such as DUNE will also make significant demands on the HEP data processing infrastructure. It is therefore essential that event reconstruction 
algorithms and software triggers continue to evolve so that they are able to efficiently exploit future computing architectures and deal with this increase in data rates without loss of physics capability. 

We have identified seven key areas where research and development is necessary to enable the community to exploit the full power of the enormous datasets which we will be collecting. Three of these 
areas concern the increasingly parallel and heterogeneous computing architectures which we will have to write our code for. In addition to a general effort to vectorize our codebases, we must understand 
what kinds of algorithms are best suited to what kinds of hardware architectures, develop benchmarks that allow us to compare the physics-per-dollar-per-watt performance of different algorithms across a 
range of potential architectures, and find ways to optimally utilise heterogeneous processing centres. The consequent increase in the complexity and diversity of our codebase will necessitate both a 
determined push to educate tomorrow’s physicists in modern coding practices, and a development of more sophisticated and automated quality assurance and control for our codebases. The increasing granularity 
of our detectors, and the addition of timing information to help cope with the extreme pileup conditions at the HL-LHC, will require us to both develop new kinds of reconstruction algorithms 
and to make them fast enough for use in real-time. Finally, the increased signal rates will mandate a push towards real-time analysis in many areas of HEP, in particular those with low transverse momentum 
signatures.

The success of this research and development program will be intimately linked to challenges confronted in other areas of HEP computing, most notably the development of software frameworks which are able 
to support heterogeneous parallel architectures, including the associated data structures and I/O, the development of lightweight detector models that maintain physics precision with minimal timing and 
memory consequences for the reconstruction, enabling the use of offline analysis toolkits and methods within real-time analysis, and an awareness of advances machine learning reconstruction algorithms 
being developed outside HEP and the ability to apply them to our problems. For this reason perhaps the most important task ahead of us is to maintain the community which has coalesced together in this 
CWP process, so that the work done in these sometimes disparate areas of HEP fuses coherently together into a solution to the problems facing us over the next decade.



\begin{thebibliography}{99}

\bibitem{HSF2017}
HEP Software Foundation,
\textit{A Roadmap for HEP Software and Computing R\&D for the 2020s},
HSF-CWP-2017-01, arXiv:1712.06982v2 (\href{https://arxiv.org/abs/1712.06982}{Link}).


\bibitem{Campana2016}
S. Campana,
\textit{Presentation to the 2016 Aix-les-Bains ECFA HL-LHC workshop},
Oct 3 2016 (\href{https://indico.cern.ch/event/524795/contributions/2236590/attachments/1347419/2032314/ECFA2016.pdf}{Link}).

\bibitem{Bordry2016} 
F. Bordry, 
\textit{presentation at the LHC Performance Workshop (Chamonix 2016)}, 
January 25th - 28th, 2016. (\href{http://chamonix-2016.web.cern.ch/}{Link} and \href{http://lhc-commissioning.web.cern.ch/lhc-commissioning/schedule/LHC-schedule-update.pdf}{Link})

\bibitem{LHCb2012} 
LHCb Collaboration, 
\textit{LHCb upgrade framework TDR},  
CERN-LHCC-2012-007 (2012) (\href{https://cds.cern.ch/record/1443882}{Link}).

\bibitem{ALICE2013} 
A. Kluge et al., 
\textit{Upgrade of the ALICE Readout and Trigger System}, 
CERN-LHCC-2013-019 (\href{https://cds.cern.ch/record/1603472}{Link}).

\bibitem{Zimmerman2009} 
F. Zimmerman, PoS EPS-HEP2009, 140 (2009). 
G. Apollinari et al., 
\textit{High-Luminosity Large Hadron Collider (HL-LHC): Preliminary Design Report}, 
CERN-2015-005.

\bibitem{ATLAS2015} ATLAS Collaboration, \textit{ATLAS Phase-II Upgrade Scoping Document},
CERN-LHCC-2015-020, LHCC-G-166 (2015) (\href{https://cds.cern.ch/record/2055248}{Link}).

\bibitem{CMS2015}
CMS Collaboration,
\textit{Technical Proposal for the Phase-II Upgrade of the CMS Detector}, 
CERN-LHCC-2015-010, LHCC-P-008, CMS-TDR-15-02 (2015) (\href{https://cds.cern.ch/record/2020886}{Link}). 

\bibitem{LHCb2017} 
LHCb Collaboration, 
\textit{Expression of Interest for a Phase-II LHCb Upgrade: Opportunities in flavour physics, and beyond, in the HL-LHC era},  
CERN-LHCC-2017-003 (2017) (\href{https://cds.cern.ch/record/2244311}{Link}).
  
\bibitem{Ohnishi2013} 
Y. Ohnishi, et. al., 
\textit{Accelerator design at SuperKEKB}, 
Prog. Theor. Exp. Phys. 2013 03A011 (2013) (\href{http://dx.doi.org/10.1093/ptep/pts083}{Link}). 

\bibitem{BelleII2010} 
Belle II Collaboration, 
\textit{Belle II Technical Design Report}, 
KEK Report 2010-1 (2010) (\href{https://arxiv.org/abs/1011.0352}{Link}).

\bibitem{DUNE2016} 
DUNE Collaboration, 
\textit{Long-Baseline Neutrino Facility (LBNF) and Deep Underground Neutrino Experiment (DUNE) Conceptual Design Report Volume 1: The LBNF and DUNE Projects}, 
arXiv:1601.05471 (2016) (\href{https://arxiv.org/abs/1601.05471}{Link}). 

\bibitem{Ball2014} 
A. Ball et al., 
\textit{Future Circular Collider Study Hadron Collider Parameters}, 
FCC-ACC-SPC-0001 1342402, CERN (2014) (\href{https://indico.cern.ch/event/298180/contributions/1658149/attachments/560575/772288/FCC-1401101315-DSC_HadronColliderParameters_V0.3.pdf}{Link}).

\bibitem{Ilten2016} 
P. Ilten, Y. Soreq, J. Thaler, M. Williams, and W. Xue, 
\textit{Proposed inclusive dark photon search at LHCb}, 
Phys. Rev. Lett. 116, 251803 (2016). 

\bibitem{aTLAs2017}
ATLAS Collaboration,
\textit{Trigger-object Level Analysis with the ATLAS detector at the Large Hadron Collider: summary and perspectives}, 
ATL-DAQ-PUB-2017-003 (2017) (\href{https://cds.cern.ch/record/2295739}{Link}).

\bibitem{CMS2016}
CMS Collaboration,
\textit{Search for narrow resonances in dijet final states at s√= 8 TeV with the novel CMS technique of data scouting},
Phys. Rev. Lett. 117 031802 (2016), (\href{https://cds.cern.ch/record/2149625}{Link}). 

\bibitem{LHCb2014}
LHCb Collaboration, 
\textit{LHCb Trigger and Online Upgrade Technical Design Report},
CERN-LHCC-2014-016 (2014) (\href{https://cds.cern.ch/record/1701361}{Link}).

\bibitem{Marshall2015} The Pandora Software Development Kit for Pattern Recognition
J.S. Marshall, M.A. Thomson (Cambridge U.). Jun 16, 2015. 15 pp.
Published in Eur.Phys.J. C75 (2015) no.9, 439

\bibitem{Gray2016} 
L. Gray, 
\textit{Challenges of Particle Flow reconstruction in the CMS High-Granularity Calorimeter at the High-Luminosity LHC}, 
38th International Conference on High Energy Physics (ICHEP 2016), (\href{https://indico.cern.ch/event/432527/contributions/1071751/}{Link}). 

\bibitem{Gray2017} 
L. Gray, 
\textit{4 Dimensional Trackers},
(2017) (\href{https://indico.cern.ch/event/577003/contributions/2476434}{Link}).


\bibitem{ALICE2015}
ALICE Collaboration, 
\textit{Technical Design Report for the Upgrade of the Online-Offline Computing System}, 
(\href{https://cds.cern.ch/record/2011297}{Link}).

\bibitem{Abreu2014} 
R. Abreu, 
\textit{The upgrade of the ATLAS High Level Trigger and Data Acquisition systems and their integration}, 
Proceedings of 19th IEEE-NPSS Real-Time conference 2014, pp.7097414 (2014) (\href{https://cds.cern.ch/record/1702466/}{Link}). 

\bibitem{Petrucciani2015} 
G. Petrucciani et. al., 
\textit{Mini-AOD: A New Analysis Data Format for CMS}, Journal of Physics: Conference Series 664 7 072052 (2015) 

\bibitem{Eifert2015} 
T. Eifert, et. al., 
\textit{Implementation of the ATLAS Run 2 event data model}, 
Journal of Physics: Conference Series, Volume 664, Offline software (2015) (\href{http://iopscience.iop.org/article/10.1088/1742-6596/664/7/072045}{Link}).

\bibitem{Aaij2016}
R. Aaij, et. al., 
\textit{Tesla : an application for real-time data analysis in High Energy Physics}, 
Comput. Phys. Commun. 208 (2016) 35-42 (\href{https://cds.cern.ch/record/2147693}{Link}).

\bibitem{Bird2014}
I. Bird, et. al., 
\textit{Update of the Computing Models of the WLCG and the LHC Experiments},
CERN-LHCC-2014-014, (2014), (\href{http://cds.cern.ch/record/1695401/files/LCG-TDR-002.pdf}{Link}).

\bibitem{Allton2017} 
C. Allton, et. al.,  
\textit{Computing Resources Scrutiny Group}, 
CERN–RRB–2017–056 (2017) (\href{https://indico.cern.ch/event/617821/contributions/2493341/attachments/1446478/2233022/CERN-RRB-2017-056_V2.pdf}{Link}). 

\bibitem{DUNE2015}, 
Dune Collaboration, 
\textit{LBNF/DUNE Conceptual Design Report, Annex 4B Expected Data Rates for the DUNE Detectors}, 
(2015) (\href{http://lbne2-docdb.fnal.gov/cgi-bin/ShowDocument?docid=10720}{Link}). 

\bibitem{DUNE2017}, 
Dune Collaboration, 
\textit{The Single-Phase ProtoDUNE Technical Design Report},
(2017) (\href{https://arxiv.org/abs/1706.07081}{Link}). 

\bibitem{Miyamoto2015} 
A.Miyamoto et al., 
\textit{Computing Requirements of the ILC Experiments}, 
LCC-Note (2015), (\href{http://www-jlc.kek.jp/~miyamoto/SoftwareCommonTask/docs/ILCComputing-EDMS1130485.A.1.1.pdf}{Link})

\bibitem{Richter2016} 
M. Richter, 
\textit{Online Data Compression in the ALICE O2 Facility}, 
Presented at the Computing in High Energy and Nuclear Physics conference (2016),
\href{https://indico.cern.ch/event/505613/contributions/2227264/}{Link}.

\bibitem{ATLASTwiki} 
\href{https://twiki.cern.ch/twiki/pub/AtlasPublic/TriggerOperationPublicResults/BandwidthPie2016.png}{BandwidthPie2016.png}, 
\href{https://twiki.cern.ch/twiki/pub/AtlasPublic/TriggerOperationPublicResults/Time_HLTRate_Stack_2016_07.png}{Time\_HLTRate\_Stack\_2016\_07.png}

\bibitem{CMS2014} 
CMS Collaboration, 
\textit{Description and performance of track and primary-vertex reconstruction with the CMS tracker},  
JINST 9 (2014) 10009 (\href{http://cds.cern.ch/record/1704291}{Link}). 

\bibitem{CMS2017} 
CMS Collaboration, 
\textit{Particle-flow reconstruction and global event description with the CMS detector}, 
Submitted to JINST (2017) (\href{https://arxiv.org/abs/1706.04965}{Link}).

\bibitem{LHCbRTCalib2016} 
G. Dujani and B. Storaci, 
\textit{Real-time alignment and calibration of the LHCb Detector in Run II}, 
Journal of Physics: Conference Series, Volume 664, Online computing (\href{https://cds.cern.ch/record/2017839?ln=en}{Link}).

 










\end{thebibliography}

\section{Appendix}

This section compiles short descriptions of some on-going projects within the community of relevance to the research and development roadmap identified by the working group. We provide short descriptions (typically taken from the project web page when available), and links to code and/or recent references. In some cases the code is part of a larger experiment framework, but we nevertheless felt it was important to identify these on-going projects.

We do not attempt to cover two classes of software development projects. First, we do not include projects to develop new code or to improve the performance of existing code purely within a single experiment (or group of experiments) whose code is not easily shared with other parts of HEP either because it is not publically available, because it is strictly tied to use in the processing framework of a specific experiment, or for other reasons. Such projects are numerous in nature and are of critical importance to the success of future experiments, but  typically these works can not be easily made into a commonly available toolkit.

Second are toolkits developed outside of the HEP community. Software trigger and event reconstruction algorithms leverage numerous toolkits developed outside of HEP. Such toolkits are frequently the basis of new research and development projects and are one mechanism to ensure both good community support and efficient code that is likely to evolve with computing technology. These include packages designed for linear algebra, machine learning and other mathematical libraries. It is important that the HEP community encourage the use of these toolkits for future development, but unfortunately the breath of these toolkits make them too numerous to include here. 

On-going community software projects identified by the working group:

\vskip 0.5 cm \noindent {\bf ACTS (A Common Tracking Software)}
This project is supposed to be an experiment-independent set of track reconstruction tools. The main philosophy is to provide high-level track reconstruction modules that can be used for any tracking detector. The description of the tracking detector's geometry is optimized for efficient navigation and quick extrapolation of tracks. 
\begin{itemize}
\item Project homepage: \href{https://gitlab.cern.ch/acts/a-common-tracking-sw}{GitLab homepage} 
\item References: ACTS-CDOT-Status-2017-03-07.pdf 
\end{itemize}

\vskip 0.5 cm \noindent {\bf AIDA Tracking Toolkit}
A generic, mostly framework independent, tracking toolkit. Development of this software package  is in the process of being merged with the ACTS project.
\begin{itemize}
\item Project homepage: \href{https://github.com/AIDASoft/aidaTT}{GitHub/AIDASoft/aidaTT} 
\item References: F. Gaede, et. al., “Software toolkit with tracking algorithms”, AIDA Delivery Report D2.8, (2015)  (http://cds.cern.ch/record/1982416). 
\end{itemize}

\vskip 0.5 cm \noindent {\bf Arbor}
ArborPFA is a C++ implementation of a Particle Flow Algorithm developed with the PandoraSDK framework. The idea under this clustering algorithm is based on the topological development of hadronic showers in high granularity sampling calorimeters follows an oriented-tree structure.
\begin{itemize}
\item Project homepage: \href{http://arborpfa.github.io/ArborPFA/}{GitHub homepage} 
\item References: M. Ruan and H. Videau, “Arbor, a new approach of the Particle Flow Algorithm”, in Proceedings, International Conference on Calorimetry for the High Energy Frontier (CHEF2013), p. 316. (2013) (https://arxiv.org/abs/1403.4784).
\end{itemize}

\vskip 0.5 cm \noindent {\bf Cross architecture Kalman Filter}
The aim of this project is to produce a fast and efficient Kalman Filter, while preserving correctness of results, across a variety of architectures.
\begin{itemize}
\item Project homepage: \href{https://gitlab.cern.ch/dcampora/cross_kalman}{GitLab/dcampora/cross\_kalman} 
\item References: D. H. C. Perez, Presentation at CHEP 2016: https://indico.cern.ch/event/505613/contributions/2227256/ 
\end{itemize}

\vskip 0.5 cm \noindent {\bf FastJet}
A software package for jet finding in pp and e+e− collisions. It includes fast native implementations of many sequential recombination clustering algorithms, plugins for access to a range of cone jet finders and tools for advanced jet manipulation.
\begin{itemize}
\item Project homepage: \href{http://fastjet.fr/}{Homepage} 
\item References: M. Cacciari, G.P. Salam and G. Soyez, Eur.Phys.J. C72 (2012) 1896 [arXiv:1111.6097].
\end{itemize}

\vskip 0.5 cm \noindent {\bf HEP.TrkX}
Project to evaluate and broaden the range of computational techniques and algorithms utilized in addressing HEP tracking challenges. Specifically the project will provide a framework to develop and evaluate new algorithms for track finding and classification, that will be demonstrated by applying advanced pattern recognition techniques to track candidate formation. On-going research includes deep neural networks applied to HL-LHC online and offline tracking.
\begin{itemize}
\item Project homepage: \href{https://heptrkx.github.io/}{GitHub} 
\item References: S. Farrell, et. al. “The HEP.TrkX project”, Presentation at the Connecting the Dots workshop, Orsay 2017 (Farrell\_HEPTrkX\_CTD2017.pdf). 
\end{itemize}

\vskip 0.5 cm \noindent {\bf Kalman-Filter tracking on parallel architectures}
This project aims to develop tracking algorithms based on the Kalman Filter for use in a collider experiment that are fully vectorized and parallelized. These will be usable with parallel processor architectures such as Intel's Xeon Phi and GPUs, but yet maintain and extend the physics performance required for the challenges for the High Luminosity LHC (HL-LHC) planned for the 2020s.
\begin{itemize}
\item Project homepage: \href{http://trackreco.github.io}{GitHub} 
\item References: Parallelized Kalman-Filter-Based Reconstruction of Particle Tracks on Many-Core Processors and GPUs - Submitted to proceedings of Connecting The Dots / Intelligent Trackers 2017 (Orsay) arXiv:1705.02876. Kalman filter tracking on parallel architectures - Proceedings of the 22nd International Conference on Computing in High Energy and Nuclear Physics (CHEP 2016) (San Francisco) arXiv:1702.06359.
\end{itemize}

\vskip 0.5 cm \noindent {\bf PandoraPFA}
Toolkit of particle flow algorithms and a framework for developing particle flow based reconstruction approaches.
\begin{itemize}
\item Project homepage: \href{https://github.com/PandoraPFA}{GitHub/PandoraPFA} 
\item References: M.A. Thomson, Particle Flow Calorimetry and the PandoraPFA Algorithm, Nucl. Instr. Meth. Phys. Res. A 611 (2009) 25; arXiv:0907.3577. 
\end{itemize}

\vskip 0.5 cm \noindent {\bf Pixel Tracking on GPUs} 
Fast and parallelizable algorithms for track seeding (in particular for Cellular Automation algorithm).
\begin{itemize}
\item Project homepage: Currently part of \href{https://github.com/cms-sw/cmssw}{GitHub/cms-sw/cmssw}. To be integrated into ACTS project.
\item References: \href{https://indico.cern.ch/event/567550/contributions/2627138/attachments/1512745/2359625/201708_Felice_ACAT17.pptx}{ACAT presentation} 
\end{itemize}

\vskip 0.5 cm \noindent {\bf PODIO}
C++ library to support the creation and handling of data models in particle physics. It is based on the idea of employing plain-old-data (POD) data structures wherever possible, while avoiding deep-object hierarchies and virtual inheritance. This is to both improve runtime performance and simplify the implementation of persistency services.
\begin{itemize}
\item Project homepage: \href{https://github.com/hegner/podio}{GitHub/hegner/podio} 
\item References: B. Hegner and F. Gaede, “PODIO: Design Document for the PODIO Event Data Model Toolkit”, AIDA-2020-NOTE-2016-004 (2016) (https://cds.cern.ch/record/2212785). 
\end{itemize}

\end{document}